\documentclass[entropy,article,submit,moreauthors]{mdpi} 
\usepackage{bm}

\firstpage{1} 
\pubvolume{1}
\issuenum{1}
\articlenumber{0}
\pubyear{2021}
\copyrightyear{2020}
\datereceived{} 
\dateaccepted{} 
\datepublished{} 
\hreflink{https://doi.org/} 

\Title{Objective quantum fields, retrocausality and ontology}

\TitleCitation{Objective quantum fields, retrocausality and ontology}

\Author{Peter D Drummond\orcidA{} and  Margaret D Reid\orcidB{}}

\AuthorNames{Peter D Drummond and Margaret D Reid}

\AuthorCitation{Drummond, Peter D  and Reid, Margaret D }
\address{Swinburne University of Technology, Melbourne 3122, Australia.}

\corres{Correspondence: peterddrummond@gmail.com}

\abstract{
We compare different approaches to quantum ontology. In particular,
we discuss an interpretation of quantum mechanics that we call objective
quantum field theory (OQFT), which involves retrocausal fields.
Here, objective implies the existence of fields independent of an
observer, but not that the results of conjugate measurements are predetermined:
the theory is contextual. The ideas and analyses of Einstein and Bohr
through to more recent approaches to objective realism are discussed.
We briefly describe measurement induced projections, the guided wave
interpretation, many-universes, consistent histories, and modal theories.
These earlier interpretations are compared with OQFT. We argue that
this approach is compatible both with Bohr's quantum complementarity,
and Einstein's objective realism.
}


\begin{document}

\section{Introduction}

There are few issues in science more important than the problem of
``what is''. This question is at the heart of philosophy since Plato's
cave analogy \citep{jowett1888republic}, and it has led to scientific
breakthroughs, like Rutherford's discovery of the nucleus \citep{rutherford1911lxxix}.
Such questions have been common in quantum mechanics since the early
discussions of Bohr \citep{bohr1987essays}, Einstein \citep{Einstein1935}
and Schr\"odinger \citep{schrodinger1935gegenwartige}. The fundamental
issue is understanding the nature of reality.

Today, there are many viewpoints on this question among physicists.
Recent analyses present the case that the wave-function has a real
existence \citep{pusey2012reality}. There is also the notion that
it simply carries information \citep{fuchs2014introduction}. An old
argument against wave-function or ``quantum state'' reality is that
the standard Copenhagen approach implies that the wave-function ``collapses''
on measurement \citep{dirac1981principles}. This has often been thought
to imply a lack of wave-function reality.

Here we explore another viewpoint, based on time-symmetry, and give
arguments in favor of it. A quantum model has been proposed in which
there are objective fields, with a probability \citep{drummond2020retrocausal}
based on the Q-function of quantum mechanics \citep{Husimi1940,joseph2018phase}.
In this approach the quantum state is a computational aid, which one
may wish to use, or not. The fields are defined to exist in a Lorenz-invariant
space-time manifold. Their probability is determined by a probabilistic
action principle, without requiring imaginary time \citep{drummond2021time}.
The important feature is that the theory is time-symmetric and involves
retrocausality.

One can also ask ``why even worry about 'what is'?''. This is much
easier to explain. Max Born asked in his Nobel lecture about quantum
mechanics, ``\emph{what is the reality which our theory has been
invented to describe?}'' \citep{born1955statistical}. Physicists
have always asked about deeper realities. This has been highly productive,
whether investigating what lies inside the atom, or outside the solar
system. In this case, the question is: what reality is there beneath
the abstract formalism of quantum mechanics?

The argument against the wave-function being a complete description
of reality is straightforward. Measurement is ill-defined, in the
sense that whatever a measurement does is no different in principle
to any other interaction. There is no ``collapse'' in normal quantum
time-evolution. The idea that collapse occurs on measurement is often
taken to imply that the wave-function is not objectively real. However,
it could describe a state of knowledge that changes on measurement.
This does not imply anything strange, just acquisition of knowledge.

If a retrocausal approach seems too simple, we might answer: why not?
An objective model of fields can be relativistic. Retrocausal dynamics
has been studied in physics many times since the work of Dirac, Feynman
and Wheeler on electrodynamics \citep{dirac1938classical,wheeler1945interaction}.
It has also been used to understand Bell violations \citep{pegg1980objective,cramer1980generalized}.
It could be that it explains quantum measurement randomness. This
appears consistent with the vacuum noise present in OQFT. 

Many no-go theorems in physics are based inherently on the assumption
of the present affecting the future, but they disallow the reverse,
which often means that they implicitly assume non-contextuality \citep{harrigan2010einstein}.
Other no-go theorems \citep{arntzenius1994spacelike,maudlin1996space}
have ruled out retrocausal explanations of Bell non-locality for classical
noncyclic causal processes, or point to fine-tuning problems \citep{wood2015lesson,Allen2021QuantumPhysRevX.7.031021}.
None of these rule out an objective, contextual theory with cyclic
loops \citep{berkovitz2002causal}, as proposed here. Since classical
causal relations are negated by quantum states, more general objective
theories are required. 

A retrocausal approach has the virtue of being conceptually simple
compared to some alternatives, like many-worlds theories \citep{dewitt1973many}.
It has the great advantage that measurement is treated on the same
basis as other processes. There is no special treatment of measurement,
with a non-unitary collapse. Measurements involve the same physical
laws as anything else. Another advantage is to give a description
consistent with relativity, that does not require violation of special
relativity with superluminal disturbances, as in early guiding-wave
theories \citep{BohmPhysRev.85.166}. Scientific theories are always
provisional, so this proposal of an alternative is an existence theorem,
demonstrating that modeling quantum reality with physical field configurations
in space-time is not impossible. 

The purpose of this paper, in summary, is to discuss a novel interpretation
of quantum mechanics as describing reality via a model of objective
quantum fields. For simplicity, we do not treat objective collapse
theories \citep{Pearle1976PhysRevD.13.857,ghirardi1986unified}. These
inherently lead to a different set of predictions from quantum mechanics.
Such theories also include an asymmetry under time-reversal, and are
different to the approach that we outline here. Our focus in this
paper is mostly on concepts. We hope that this may be useful for those
wishing to obtain a general picture of the philosophy behind the approach,
without mathematical details that are derived elsewhere \citep{drummond2020retrocausal,drummond2021time,Reid2021}.

\section{Objective fields}

We now explain how the proposed interpretation of quantum mechanics,
objective quantum field theory (OQFT), unifies most of the views of
Einstein and Bohr. Our proposal is presented in \citep{drummond2020retrocausal},
and is motivated by the Q-function phase space representation. As
such, it is an objective model of fields in space and time. It represents
the known physics of quantum fields, our most fundamental theory.
This approach is compatible with the standard model, is consistent,
and appears empirically viable. It includes random vacuum noise, in
accord with experiments that have random results. The observation
of eigenvalues, and the violation of Bell inequalities can both be
replicated \citep{drummond2020retrocausal,Reid2021}.

However, the theory must necessarily include a physical model of the
measuring instrument to be a complete picture of measurement. This
provides a natural, operational way to explain contextual results.
The model is different to the earlier interpretations described below.
It proposes that there are many potential universes, but not that
they all exist \citep{dewitt1973many}. One can imagine having six
possible outcomes from throwing a dice. However, they do not all exist
simultaneously.

Nor does decoherence play a fundamental role. In OQFT, what is most
important for measurement is amplification \citep{reid2017interpreting}.
This makes microscopic quantum events large enough for our macroscopic
neurons to register them. Decoherence occurs in as much as macroscopic
signals couple strongly to their environment. This is a consequence,
not a cause of measurement.

The model is described by a probability at time $t$, $P(\bm{\phi},t)$,
for a set of fields $\bm{\phi}\left(r\right)$ that are objectively
real. These exist in a four-dimensional space-time with a coordinate
$r=\left(t,\bm{r}\right)$. This includes the fact that there are
matter-waves. The fields can diffract and interfere, and have quantum
statistics. There are no point-like particles, although fields may
be localized due to interactions. This is the difference between using
fields, and the particle ontology of Bohm \citep{BohmPhysRev.85.166}.
In the OQFT approach, one defines a vector field amplitude $\bm{\phi}$,
and a coherent state projector $\hat{\Lambda}\left(\bm{\phi}\right)$.
The probability of $\bm{\phi}$ for a quantum density matrix $\hat{\rho}\left(t\right)$
is a generalization of the Husimi Q-function \citep{Husimi1940},
defined as
\begin{equation}
P(\bm{\phi},t)=Tr\left(\hat{\Lambda}\left(\bm{\phi}\right)\hat{\rho}\left(t\right)\right),\label{eq:Q}
\end{equation}
 where:
\begin{equation}
\int P\left(\bm{\phi},t\right)d\bm{\phi}=1.
\end{equation}
Here $P(\bm{\phi},t)$ is the marginal probability at time $t$, relative
to $\mathcal{P}\left[\bm{\phi}\right]$, which is the full space-time
functional probability. The generalized phase-space coordinate $\bm{\phi}$
includes both Bose and Fermi fields, and $\hat{\Lambda}\left(\bm{\phi}\right)$
is an outer product of bosonic and fermionic projectors. For example,
in the case of a complex scalar bosonic field, as in the Higgs-Englert-Guralnik
model \citep{englert1964broken,higgs1964broken,guralnik1964global},
the objective quantum field for $M$ discrete modes in a volume $V$
is a complex field representing an operator field $\hat{\phi}$: 
\begin{equation}
\phi\left(r\right)=\sum_{n}\frac{1}{\sqrt{2E_{k}V}}\left[e^{i\bm{k}_{n}\cdot\bm{r}}\alpha_{n}\left(t\right)+e^{-i\bm{k}_{n}\cdot\bm{r}}\beta_{n}^{*}\left(t\right)\right].
\end{equation}
The $\bm{k}_{n}$ are field momenta, $\bm{\alpha}$ and $\bm{\beta}$
are coherent state vectors for particles $\left(+\right)$ and antiparticles
$\left(-\right)$, and the projector $\hat{\Lambda}_{b}\left(\bm{\phi}\right)$
for bosonic coherent states $\left|\bm{\alpha}\right\rangle _{+},\left|\bm{\beta}\right\rangle _{-}$
is: 
\begin{equation}
\hat{\Lambda}_{b}\left(\bm{\phi}\right)=\frac{1}{\pi^{2M}}\left|\bm{\alpha}\right\rangle \left\langle \bm{\alpha}\right|_{+}\left|\bm{\beta}\right\rangle \left\langle \bm{\beta}\right|_{-}.
\end{equation}

As usual, one can take infinite mode limits as $M,V\rightarrow\infty$,
provided the Hamiltonian is renormalizable. Fermionic coherent states
require a generalized Gaussian projection operator $\hat{\Lambda}_{f}$,
which is a function of an antisymmetric real matrix that describes
Fermi field correlations \citep{Corney_PD_JPA_2006_GR_fermions,Corney_PD_PRB_2006_GPSR_fermions,Corney_PD_PRL2004_GQMC_ferm_bos,ResUnityFGO:2013,joseph2018phase}.
The matrix, after Fourier transforming, represents a product of two
operator Majorana fields, $i\hat{M}_{i}\left(\bm{r}_{1}\right)\hat{M}_{j}\left(\bm{r}_{2}\right)$.

Objective fields have the hallmark of elements of reality. There is
a positive probability density of fields $\bm{\phi}$ occurring at
a given time, and the total probability is unity. Thus, normal laws
of probability hold. Hence, it is not unreasonable to claim that $\bm{\phi}$
exists ontologically, although $P(\bm{\phi},t)$ is epistemic, or
knowledge-based. Since $P$ is equivalent to a wave-function, this
implies that the wave-function is also epistemic. It describes knowledge
via probability, and therefore is \emph{not} ontological. This is
not inconsistent with the phase-space coordinate $\bm{\phi}$ being
viewed as real.

The task of a physicist is to determine the probabilities that certain
trajectories exist. This creates an epistemic distribution of trajectories
equivalent to a wave-function. Despite this, a measurement doesn't
change the universe's trajectory except according to the fundamental
action principle. There is a self-consistency requirement, which is
not surprising. Cognition has no effect apart from biochemistry, unless
it results in the observer's actions, but this too is part of the
universe.

There is an increase of knowledge on measurement, if one wishes to
take cognition into account. The information itself exists relative
to an observer, their observations, and their deductions. The objective
universe is not greatly changed when an observer determines what they
know about their apparatus. There are neurological changes, described
by the same laws of physics as everything else. The universe would
evolve in a similar way if there were no observers. The moon exists
even when unobserved.

More details are given elsewhere \citep{drummond2020retrocausal,drummond2021time,Reid2021},
although this is still very much work in progress. Not all the calculations
of quantum field theory have been explicitly repeated. Certainly,
much remains to be calculated. Conventional quantum mechanics, following
Dirac's description of the Copenhagen approach, is a highly successful
theory, with over a century of complex theoretical work behind it.
The approach described here is not intended to replace this work.
However, it may make computations easier in future, as new techniques
are developed.

In the calculations carried out so far \citep{drummond2020retrocausal,Reid2021},
no inconsistencies with standard QFT were found \emph{provided a model
of the measuring instrument is included}. In other words, OQFT gives
empirical results similar to those obtained from standard QFT, if
one only takes observational account of macroscopic outputs of measuring
instruments. This general approach was emphasized in Bohr's early
writings \citep{Bohr1928Quantum,Bohr1935CanQuant}. State preparation
has rather similar properties. That is, a realistic model of quantum
state preparation would require essentially a time-reversal of the
measurement argument, with inputs from some operational, macroscopic
device. Neither case requires the field coordinate $\bm{\phi}$ to
have a one-to-one correspondence with a traditional quantum state
$\left|\psi\right\rangle $. 

Detailed results were given for the measurement dynamics of the phase
coordinates $x$ and $p$, for a single mode field prepared in an
eigenstate $|x_{i}\rangle$ of the $X$ field quadrature \citep{drummond2020retrocausal,Reid2021}.
As is well known, the Husimi Q function for an eigenstate $|x_{i}\rangle$
is a Gaussian with a variance (noise $\eta$) of order $\hbar$ in
the phase space variable $x$. Noise is present, even though this
is an eigenstate. Solving for the dynamics of the measurement modeled
as a parametric amplifier reveals retrocausal trajectories for $x$.
The result is that the measurement gives a distinct amplified outcome
$gx_{i}$, where $g$ is the amplifier gain, with minimal noise to
signal ratio. This allows the eigenvalue $x_{i}$ to be inferred from
the macroscopic signal, since the noise $\eta$ is \emph{not} amplified. 

Other analyses have recently obtained similar conclusions \citep{Friederich2021Introducing}.
Hence, measurements will lead to a well-defined spectrum of outcomes
\citep{Reid2021}. In this model, measurement is the process that
gives a distinct eigenvalue. Including a measuring instrument is the
natural source of contextual measurement results. This is important,
and was an issue also emphasized very strongly by Bohr. With an appropriate
model of an amplifier present, there is no need to carry out projections.
In fact, these are not meaningful in an objective model. This approach
also provides a source of the random noise observed in measurements,
an issue often neglected in realistic models.

We note that the OQFT model is consistent with macroscopic realism,
if that concept is carefully defined. This is because the system in
a superposition of macroscopically distinguishable eigenstates $|x_{1}\rangle$
and $|x_{2}\rangle$ may be interpreted to be probabilistically in
one or other state that gives a definite \emph{value}, $x_{1}$ or
$x_{2}$, for the result of the measurement $X$ \citep{Reid2021,drummond2020retrocausal,reid2017interpreting}.
There is thus a consistency with macroscopic realism, without the
need to introduce decoherence. While this appears to contradict claims
in the literature of violations of macroscopic Bell inequalities \citep{thenabadu2020testing}
and Leggett-Garg's macro-realism \citep{leggett1985quantum,emary2013leggett},
this can be explained, once one accounts for the dynamical nature
of the measurement process \citep{thenabadu2020bipartite}. Of course,
the consistency with macroscopic realism as considered here does not
imply the system in the superposition was actually in one or other
\emph{eigenstates} $|x_{1}\rangle$ or $|x_{2}\rangle$, prior to
measurement. This would be a much stronger definition of macroscopic
realism which is known to be falsifiable, since macroscopic superpositions
and mixtures can be experimentally distinguished, although with some
difficulty \citep{yurke1986generating}.

The distinct values of $\bm{\phi}$ we use do not correspond to orthogonal
projections. The projectors are determined by the symmetric spaces
of Lie groups \citep{cartan1926classe,cartan1935domaines,perelomov1972coherent,Arecchi1972,Gilmore1975,Zhang1990}
corresponding to the field commutators. This is a mathematical criterion,
unrelated to any measurement. A phase-space coordinate $\bm{\phi}$
evolves in time to give a phase-space trajectory $\bm{\phi}(t)$.
The trajectory $\bm{\phi}(t)$ is a set of classical fields in space-time,
which exists at all space-time events. For bosons it can be written
as $\bm{\phi}\left(t,\bm{x}\right)$. In the case of fermions, one
has to include field correlations.

\section{Causality and retrocausality}

The idea of future boundary conditions in electrodynamics was proposed
over a century ago as a field-free variational action principle with
direct action at a distance \citep{schwarzschild1903elektrodynamik,tetrode1922causal,fokker1929invarianter}.
Dirac \citep{dirac1938classical} introduced a theory of radiation-reaction
in which boundary conditions in the future were used. This approach
was formalized by Wheeler and Feynman, who called it absorber theory
\citep{wheeler1945interaction}. A quantum absorber theory was later
developed \citep{hoyle1971electrodynamics}, equivalent to quantum
electrodynamics. The application of absorber theory to Bell violations
\citep{cramer1980generalized,pegg1980objective} was used to explain
the apparent lack of causality in Bell violations, which led to transactional
quantum theory \citep{cramer1986transactional}.

Criticisms were leveled at this approach \citep{arntzenius1994spacelike,maudlin1996space},
with the claim that retrocausality might lead to contradictions. In
response, it was argued that such no-go theorems relied on untenable
assumptions about probabilities in causal loops \citep{berkovitz2002causal,kastner2006cramer}.
More recent reviews \citep{price2008toy,wharton2020colloquium} have
treated toy models of retrocausal quantum fields, and discuss these
issues in more detail.

Objections have been raised to the assumption that realistic models
must not be contextual, where: '\emph{one wishes some aspect of \textquotedbl reality\textquotedbl{}
to be describable in a manner that is independent of the act of experimentation}'
\citep{Shrapnel2018causationdoesnot}. This raises the question of
why one should wish this, since making this assumption eliminates
potential models, including OQFT (unless we claim that the ``aspect''
is reality at a macroscopic level). Given that the \emph{'act of experimentation'}
is part of the universe, it is not clear why reality has to be independent
of it. 

We will discuss how one might resolve these questions using the objective
model of real fields described in the previous section. In this model,
the fields are regarded as having an objective existence, with retrocausal
dynamics. We now outline a proof of equivalence to the standard model
of quantum field theory. The time evolution of quantum fields in quantum
field theory (QFT) is conventionally described by the Dirac-Feynman
quantum action principle, which has an imaginary exponent. This is
equivalent to the Schr\"odinger equation,
\begin{equation}
\frac{d\hat{\rho}\left(t\right)}{dt}=\frac{i}{\hbar}\left[\hat{\rho}\left(t\right),\hat{H}\right].\label{eq:Schrod}
\end{equation}
Eq (\ref{eq:Q}) and (\ref{eq:Schrod}) make it seem that $P(\bm{\phi},t)$
relies on a wave-function or density matrix $\hat{\rho}\left(t\right)$
evolving in time. However, this is not required. By using appropriate
identities, one can obtain a partial differential equation, in the
fields alone, of form:
\begin{equation}
\frac{dP(\bm{\phi},t)}{dt}=\mathcal{D}P(\bm{\phi},t).\label{eq:GFPE}
\end{equation}

If the quantum Hamiltonian $\hat{H}$ has nonlinearities that are
at most quartic in the quantum fields, $\mathcal{D}$ has up to second
order partial derivatives in the fields $\bm{\phi}.$ This is an algebraic
property of the differential identities used to transform Eq (\ref{eq:Schrod})
into the generalized Fokker-Planck equation, (\ref{eq:GFPE}).

The resulting partial functional differential equation is similar
to a Fokker-Planck equation, describing diffusion in a noisy environment,
except that the diffusion matrix is not positive-definite, and has
a zero trace. As a result, the trajectories obey a real action principle
\citep{stratonovich1971probability,Graham1977Covariant,graham1977lagrangian}
of an unusual type \citep{drummond2020retrocausal,drummond2021time}.
The probability functional of a complete space-time trajectory $\mathcal{P}\left[\bm{\phi}\right]$
is expressed in terms of a time-symmetric Lagrangian $L_{\phi}\left(\bm{\phi},\dot{\bm{\phi}}\right)$,
different to the conventional Lagrangian of mechanics. We suppress
the space-dependence of fields in the following, for brevity. The
path integral satisfies mixed (Robin-type) boundary conditions in
the past and future:
\begin{equation}
\mathcal{P}\left[\bm{\phi}\right]=\exp-\int L_{\phi}\left(\bm{\phi}(t),\dot{\bm{\phi}}(t)\right)dt\,.
\end{equation}

In summary, time-evolution of the objective fields is described by
an action principle which determines the probability of a particular
trajectory $\bm{\phi}\left(r\right)$ occurring, given boundary conditions
in the past and the future. The probability of any particular evolution
is defined according to a path integral. Unlike the Dirac-Feynman
quantum action principle, the action is real. It has mixed boundary
conditions both in the past and in the future. These fields are therefore
neither the elements of reality of Einstein, Podolsky and Rosen \citep{Einstein1935},
nor hidden variables in the sense of Bell \citep{Bell1964}.

Unlike the Dirac-Feynman path integral, this stochastic path integral
cannot be used to quantize all interactions. If the Hamiltonian has
higher than quartic nonlinearities, Eq (\ref{eq:GFPE}) has third
or higher order differential terms. In such cases, no path-integral
result is known. In the similar case of a forward Fokker-Planck equation,
it is known that such equations have no probabilistic interpretation
\citep{pawula1967approximation}.

In the OQFT approach, interacting fields are divided into complementary
quadratures so that 
\begin{equation}
\bm{\phi}=\left[\bm{\phi}^{+},\bm{\phi}^{-}\right],
\end{equation}
where there are two quadratures $\bm{\phi}^{\pm}$. One propagates
forward in time, one backward in time, so that the result of applying
the action principle is a forward-backward stochastic differential
equation, of the general form: 
\begin{align}
\bm{\phi}^{+}\left(t\right) & =\Phi^{+}\left(\bm{\phi}^{-}\left(0\right)\right)+\int_{0}^{t}\bm{A}^{+}\left[\bm{\phi}\left(t'\right)\right]dt'+\int_{0}^{t}d\bm{w}^{+}\left(t'\right)\nonumber \\
\bm{\phi}^{-}\left(t\right) & =\Phi^{-}\left(\bm{\phi}^{+}\left(T\right)\right)-\int_{t}^{T}\bm{A}^{-}\left[\bm{\phi}\left(t'\right)\right]dt'-\int_{t}^{T}d\bm{w}^{-}\left(t'\right)\label{eq:retrocausal}
\end{align}

Here $\bm{w}^{\pm}$ are stochastic fields, and $\bm{A}^{\pm}$ represent
classical-like drift functionals. In the initial and final boundary
conditions $\Phi^{\pm}$, $\bm{\phi}^{-}\left(T\right)$ may depend
on $\bm{\phi}^{+}\left(T\right)$, and/or $\bm{\phi}^{+}\left(0\right)$
may depend on $\bm{\phi}^{-}\left(0\right)$. This gives an explicit
cyclic dependence between the fields in addition to those present
in the forward-backwards equations due to cross-coupling. The combination
of retrocausality and cyclic dependency is the difference between
an objective field theory and conventional causal quantum mechanics.

The entire field $\bm{\phi}(t,\bm{x})$ is the basic element of reality
here. This can be transformed by relativistic frame transformations,
so it is meaningful in any frame of reference, and is a relativistic
invariant. The value of $\bm{\phi}$ at a fixed time is not a unique
ontological object. This would obviously be frame-dependent, since
it depends both on the chosen frame of reference, and on a particular
choice of time coordinate. The full trajectory $\bm{\phi}\left(t,\bm{x}\right)$
does not have this limitation of frame-dependence.

We cannot impose a unique time coordinate on a physical system in
an observer-independent way. One therefore should not compare the
objective field theory ontology with hidden-variable theories \citep{Bell1964},
which propose that reality is described by a set of variables defined
at a single time-like surface. Equally importantly, hidden-variables
are assumed to be independent of any future events.

There are limitations to these methods. Objective field trajectories
\emph{only} exist for Hamiltonians at most quartic in the field operators,
which includes quantum electrodynamics \citep{Tomonaga1946Relativistic,Feynman1948RelativisticPhysRev.74.1430,Schwinger1948QuantumPhysRev.74.1439,Schwinger1949QuantumPhysRev.75.651},
quantum chromodynamics \citep{thooft1971renormalizable,tHooftveltman1972regularization}
and the Higgs-Englert-Guralnik model \citep{englert1964broken,higgs1964broken,guralnik1964global}:
the standard model of physics. This covers all that is empirically
known in quantum theory. The OQFT approach is able to treat the part
of quantum field theory that is physically relevant. Higher-order
nonlinearities do not have an OQFT description, and are not renormalizable.
They are not considered fundamental in standard quantum field theory
\citep{kibble1980some,zee2010quantum}, although for different reasons.

By contrast, conventional Lagrangian methods \citep{dirac1981principles}
may include \emph{any} nonlinear coupling, almost all of which are
unphysical. Thus, OQFT is a more restricted theory than standard QFT.
The limitation to quartic nonlinearities means that the OQFT approach
is not applicable to most first quantized theories, which often use
non-polynomial interaction potentials. An example is the Coulomb interaction.
This is not a serious problem, because first quantization is not a
fully correct physical theory.

Using more limited techniques might seem unproductive. We argue that
the opposite is true. The 'consensus' theory of epicycles in geocentric
Ptolemaic astronomy was very flexible, but treated the earth as the
center of the universe. The Copernican approach of Newton, although
restricted, was more scientifically useful \citep{kuhn1957copernican},
and eliminated the earth-bound observer as being central. The present
approach also removes an anthropomorphic bias, namely that of the
Copenhagen observer. If OQFT is more restricted, it has the merit
of being more easily falsified, and therefore complies with what is
needed in a scientific theory \citep{popper1989logik}.

Quantization of gravity is outside the scope of this article. Previous
attempts at quantizing gravity are problematic. One issue is a lack
of renormalizability \citep{gomis1996nonrenormalizable,zee2010quantum},
with infinities that can't be removed. Another issue is how to treat
time \citep{isham1993canonical}. Covariance under frame transformations
is essential in general relativity. This is inconsistent with the
global role played by time in conventional quantum measurement theory.
Such problems appear absent in OQFT, although further investigation
is needed.

\section{The EPR paradox}

Bohr and Einstein were pioneers in the earliest years of quantum theory,
and engaged in several famous debates \citep{bohr1996discussion}.
It is often claimed that Bohr won these debates. Certainly the Copenhagen
school that Bohr founded was very influential. Yet Einstein was also
important, in pointing out the subtle paradoxes that resulted from
this approach \citep{Einstein1935,Reid:2009_RMP81}, and questioning
the completeness of the wave-function description. These issues are
still debated today.

Their argument led to a draw. Both Einstein and Bohr were correct
in their own way. Each had important ideas that are vital to the model
proposed here. Neither was truly wrong, just lacking the modern mathematical
tools and ideas that are available today. This does not in any way
detract from the success of the Copenhagen school, nor from the significance
of Einstein's ideas about space and time.

It is straightforward to summarize Einstein's views. As a communicator
he was prolific and skillful. His main conceptual guidelines are described
in his autobiographical notes \citep{einstein1949albert}. Diecks
\citep{dieks1988formalism} has summarized Einstein's essential principles
for a physical theory as follows:
\begin{enumerate}
\item \emph{``The theory should be objective, in the sense of observer
independent;}
\item \emph{The description should be about fields and/or particles located
in space and time;}
\item \emph{Everything that exists should have its counterpart in the theoretical
description''.}
\end{enumerate}
In work on the possible incompleteness of quantum mechanics, Einstein
also introduced classically causal ``elements of physical reality'',
which are not thought to exist today \citep{Einstein1935}.

Bohr's rhetoric is sometimes criticized, but it rewards careful reading.
His main philosophy \citep{Bohr1928Quantum} was that: \emph{``this
(quantum) postulate implies a renunciation as regards the causal space-time
co-ordination of atomic processes}'', and ``\emph{any observation
of atomic phenomena will involve an interaction with the agency of
observation not to be neglected}'' His riposte \citep{Bohr1935CanQuant}
to Einstein's claim about the incompleteness of quantum theory, repeated
his point that the operational procedures of measurement are a vital
part of any correct description of physical reality. This is surely
not that unreasonable. To test a theory, one must do experiments,
using instruments that are also part of any external objective reality.
It should not be very surprising if reality is affected by these measurements.

This is not inconsistent with Einstein's three principles enumerated
above. The fact that a measuring device can disturb what it measures
does not contradict the idea that there is objective behavior in the
universe, nor does this require an observer. Bohr summarized this
in his response to the Einstein, Podolsky and Rosen paper \citep{Bohr1935CanQuant}:
``\emph{In fact this new feature of natural philosophy means a radical
revision of our attitude as regards physical reality}''.

We agree both with the need for treating the meter, and with a renunciation
of traditional space-time causality. Similarly, the need for invariance
under a change in reference frame was vital to Einstein's relativity
theory, and also must be incorporated in any modern physical theory.

Bohr's general statement was later summarized by others as implying
a need to collapse the wave-function when describing measurements
\citep{dirac1981principles}. This extension of Bohr's ideas was not
proposed by Bohr himself. It is useful in obtaining predictions, but
conceptually it is the main drawback of the approach. It is unclear
how to define a measurement in the first place, and collapse is attributed
by some to a type of mental process \citep{Stapp2001Quantum}. This
makes it questionable that a wave-function is physically real. How
can an ontological model be changed so dramatically given such a poorly
defined process?

Einstein's third condition is often ignored, yet it is worth careful
examination. When a measurement takes place, we are familiar with
the idea that a system in an eigenstate yields an eigenvalue, which
is regarded as existing beforehand, and is present in the theory.
More typically, the measurement yields a random value which \emph{cannot}
be obtained from the wave-function alone. This does not satisfy Einstein's
criteria. Essentially, Born excused this failing by calling quantum
mechanics a statistical theory, but he also recognized that it left
an open question \citep{born1955statistical}: a statistical theory
of \emph{what?}

This open question is what the objective field approach attempts to
answer.

Between them, Bohm and Bell took up the challenge of Einstein and
Bohr, each trying to establish their understanding of ``what is''.
They were less adversarial than Einstein and Bohr, however. Bohm,
extending earlier ideas of de Broglie, created a model of reality
\citep{BohmPhysRev.85.166} to disprove a claim by von Neumann that
realistic models were not possible \citep{von2018mathematical,bub2010neumann}.
The claim of von Neumann was only valid in regard to a limited class
of models. Bohm's theory described non-relativistic particles moving
in a potential created by the non-relativistic Schr\"odinger wave-function,
which is outside this class of theories. Yet it hardly seems to fit
into a modern picture of quantum fields. Although relativistic extensions
have been proposed, they appear incomplete \citep{deckert2019persistent}.

The objective model described by Bohm does have drawbacks. It is nonlocal,
with action at a distance between the particles. It also relies on
solving for the wave-function evolution. On top of this, it adds real
particles as a superstructure. This seems to lack a coherent picture,
as one cannot easily obtain a simple, unified description. However,
this theory does make it clear that a realistic model of quantum mechanics
is not impossible, an important breakthrough given von Neuman's earlier
claim.

Bell's contribution was to demonstrate which models cannot work \citep{Bell1964},
and to investigate how to introduce Lorenz invariant, objective models
\citep{bell2004speakable}. He analyzed a class of models called
``local hidden variable theories''. These are a formalization taken
from Einstein, Podolsky and Rosen's work on the incompleteness of
quantum mechanics. Such theories are not the same as Bohm's, which
were inherently nonlocal. Bell was interested in understanding what
could be achieved with local theories, where there is no action at
a distance. In doing so, he introduced an additional, apparently almost
trivial assumption, that causality proceeds from the past to the future.
In other words, Bell implied that one can always compute the future
from the past\emph{,} which is also inherent in classical mechanics,
ignoring computability issues.

It is this important assumption that is discarded in the objective
quantum field model. Given that all physical laws appear to be time-reversible,
Bell's assumption formalizes an anthropomorphic prejudice that appears
inconsistent with objective reality. Without this assumption, quantum
features can co-exist with an objective physical reality. One should
not think that this gives one a license to change the past, as in
a science-fiction novel. The issue of time's arrow is a much more
macroscopic one, closely linked to questions like the growth of entropy
and the expansion of the universe.

\section{Other approaches to quantum ontology}

In this section, we give an overview of some other approaches to quantum
ontology. The purpose of this is not to give a comprehensive review.
Instead, we focus on more popular interpretations. Those that have
similarities to ours in particular, are treated. This allows us to
focus on difference that we believe are significant. The field is
developing rapidly, so not all alternative pictures can be included
here.

One current picture of reality is mostly surprising in its complexity.
Everett's original idea of a relative state interpretation \citep{Everett:1957},
did not treat objective reality per se. However, this has been later
interpreted as implying that the wave-function describes all possible
quantum universes, and that these all exist at once \citep{dewitt1973many}.
There have been many suggestions on how to identify what one of these
universes comprises. One picture \citep{zeh1970interpretation} is
that decoherence allows a particular universe to be selected.

This is not accepted by all the experts in this field. Bell, one of
the deepest thinkers on quantum foundations, was opposed to this idea
\citep{bell2004speakable}. He pointed out that decoherence does not
lead to a transition to any one universe. Instead, it simply causes
entanglement between a system and its environment.

A development of this general approach is the model of consistent
histories \citep{griffiths1984consistent,omnes1992consistent,gell1993classical}.
This attempts to obtain probabilities of histories of the universe
that have internal consistency. In order to do this, one must obtain
a set of projection operators describing the history. This is similar
in practice to the Copenhagen model, since measurement and collapse
of the wave-function is related to a projection.

In this model, one has a class operator for a history, composed of
projection operators at different times $t_{1},\ldots t_{n}$:
\begin{equation}
{\displaystyle {\hat{C}}_{H_{i}}}{\displaystyle ={\hat{P}}_{i,n_{i}}\cdots{\hat{P}}_{i,2}{\hat{P}}_{i,1}}.
\end{equation}
Central to the approach is consistency. For an initial density matrix
$\hat{\rho}$, a set of histories is termed consistent if
\begin{equation}
Tr\left[{\displaystyle {\hat{C}}_{H_{i}}}\hat{\rho}{\displaystyle {\hat{C}}_{H_{j}}^{\dagger}}\right]=0.
\end{equation}

Quoting Omnes \citep{omnes1999understanding}, there are three differences
with the Copenhagen approach:
\begin{enumerate}
\item \emph{The logical equivalence between an empirical datum... and the
result of a measurement, which is a quantum property, becomes clearer.}
\item \emph{There are two apparently distinct notions of probability in
the new approach. One is abstract and directed toward logic, whereas
the other is empirical and expresses the randomness of measurements.}
\item \emph{The main difference lies in the meaning of the reduction rule
for 'wave packet collapse'. In the new approach, the rule is valid
but no specific effect on the measured object can be held responsible
for it. Decoherence in the measuring device is enough.}
\end{enumerate}
Yet, it is unclear how to interpret the probabilities that are obtained
this way, nor the implication of the times $t_{1},\ldots t_{n}$.
However they are obtained, the times $t_{j}$ require a special reference
frame. A choice like this appears incompatible with special and general
relativity. In addition, the procedure may result in a set of probabilities
that are not consistent with the laws of probability theory. There
are many variations on this approach. There is also an approach called
a 'modal' theory, with similar properties \citep{fraassen1974einstein}.

A general difficulty with the multiverse, consistent histories, and
modal approaches is that there is no clear way to identify what reality
is, for example by uniquely obtaining the projectors ${\hat{P}}_{i,n_{i}}$
\citep{Dowker1995PropertiesPhysRevLett.75.3038}. One may try to link
the time-scale and projections with decoherence theory. This, however,
is not a precise formulation of reality in Born's sense. It leaves
unanswered the question of what is happening between these measurements
or projections. Nor does it have the completeness property (3) of
Einstein, concerning actual results and the randomness of measurements.

Such approaches clearly do not describe the nature of the reality
\emph{between} the projection times. Nor do they account for the origin
of the random results that are found whenever a real physical measurement
is made. Hence there are large differences between OQFT and consistent
histories. In particular, OQFT has a unique definition of what is
meant by reality. It uses projectors defined by the group-theoretic
spaces underlying the commutators. It does not employ arbitrary orthogonal
projectors, nor a sequence of operations at successive times.

There are previous attempts to introduce randomness into quantum mechanics.
We briefly mention three of them, Nelson's stochastic mechanics \citep{nelson1966derivation},
Parisi and Wu's stochastic quantization \citep{parisi1981perturbation},
and a recent phase-space based approach \citep{budiyono2017quantum}.
Nelson's approach is conceptually related to ours, although originally
just for a first quantized theory. While the idea was pioneering,
Nelson was unsure if the stochastic mechanics approach could replicate
all of quantum mechanics, let alone quantum field theory, and there
are known issues \citep{grabert1979quantum,nelson2012review} involving
repeated measurement. These problems were subsequently reanalyzed
and a resolution proposed that required using a wave-function and
projection approach, \citep{blanchard1986repeated}, which one would
argue therefore does not resolve the original measurement problem.
The Parisi and Wu approach of stochastic quantization in Euclidean
space was intended to solve gauge-fixing problems in QFT. Rotating
the path integrals to Minkowski space is not straightforward, and
may lead to negative probabilities \citep{damgaard1987stochastic}.
The approach of Budiyono and Rohrlich \citep{budiyono2017quantum}
uses both stochastic trajectories and distributions, which is similar
to the de Broglie-Bohm theory.

\section{Summary}

In conclusion, this article compares the quantum picture of reality
in an objective quantum field theory (OQFT) with that of previous
quantum physicists. We show that there are very close similarities
with the concepts held to be most important by Einstein and Bohr.
However, there are substantial differences with more recent approaches.
An OQFT has no need for multiple universes that exist simultaneously.
Wave-functions collapsing due to measurement do not occur. It places
no special importance on decoherence, nor on mental processes. New
and previously unknown physics causing decoherence is not excluded,
but also not required. The objective field approach appears empirically
adequate to describe modern quantum field theory results. Most of
the original philosophies of Einstein and Bohr are unified in this
picture.

\end{paracol}
\reftitle{References}


\begin{thebibliography}{999}

\bibitem[Jowett(1888)]{jowett1888republic}
Jowett, B.
\newblock {\em The Republic of Plato}; London; New York: Macmillan,  1888.

\bibitem[Rutherford(1911)]{rutherford1911lxxix}
Rutherford, E.
\newblock The scattering of $\alpha$ and $\beta$ particles by matter and the
  structure of the atom.
\newblock {\em The London, Edinburgh, and Dublin Philosophical Magazine and
  Journal of Science} {\bf 1911}, {\em 21},~669--688.

\bibitem[Bohr(1987)]{bohr1987essays}
Bohr, N.
\newblock {\em Essays 1958-1962 on Atomic Physics and Human Knowledge}; Number
  v. 3 in Bohr, Niels: Philosophical writings, Ox Bow Press, Woodbridge, USA,
  1987.

\bibitem[Einstein \em{et~al.}(1935)Einstein, Podolsky, and Rosen]{Einstein1935}
Einstein, A.; Podolsky, B.; Rosen, N.
\newblock {Can Quantum-Mechanical Description of Physical Reality Be Considered
  Complete?}
\newblock {\em Phys. Rev.} {\bf 1935}, {\em 47},~777--780.
\newblock
  doi:{\changeurlcolor{black}\href{https://doi.org/10.1103/PhysReview47.777}{\detokenize{10.1103/PhysReview47.777}}}.

\bibitem[Schr{\"o}dinger(1935)]{schrodinger1935gegenwartige}
Schr{\"o}dinger, E.
\newblock Die gegenw{\"a}rtige Situation in der Quantenmechanik.
\newblock {\em Naturwissenschaften} {\bf 1935}, {\em 23},~823--828.

\bibitem[Pusey \em{et~al.}(2012)Pusey, Barrett, and Rudolph]{pusey2012reality}
Pusey, M.F.; Barrett, J.; Rudolph, T.
\newblock On the reality of the quantum state.
\newblock {\em Nature Physics} {\bf 2012}, {\em 8},~475.

\bibitem[Fuchs \em{et~al.}(2014)Fuchs, Mermin, and
  Schack]{fuchs2014introduction}
Fuchs, C.A.; Mermin, N.D.; Schack, R.
\newblock An introduction to QBism with an application to the locality of
  quantum mechanics.
\newblock {\em American Journal of Physics} {\bf 2014}, {\em 82},~749--754.

\bibitem[Dirac(1981)]{dirac1981principles}
Dirac, P.A.M.
\newblock {\em The principles of quantum mechanics}; Oxford University Press,
  1981.

\bibitem[Drummond and Reid(2020)]{drummond2020retrocausal}
Drummond, P.D.; Reid, M.D.
\newblock Retrocausal model of reality for quantum fields.
\newblock {\em Physical Review Research} {\bf 2020}, {\em 2},~033266.

\bibitem[Husimi(1940)]{Husimi1940}
Husimi, K.
\newblock Some formal properties of the density matrix.
\newblock {\em Proc. Physical Math. Soc. Jpn.} {\bf 1940}, {\em 22},~264--314.

\bibitem[Joseph \em{et~al.}(2018)Joseph, Rosales-Z{\'a}rate, and
  Drummond]{joseph2018phase}
Joseph, R.R.; Rosales-Z{\'a}rate, L.E.; Drummond, P.D.
\newblock Phase space methods for Majorana fermions.
\newblock {\em Journal of Physics A: Mathematical and Theoretical} {\bf 2018},
  {\em 51},~245302.

\bibitem[Drummond(2021)]{drummond2021time}
Drummond, P.D.
\newblock Time evolution with symmetric stochastic action.
\newblock {\em Physical Review Research} {\bf 2021}, {\em 3},~013240.

\bibitem[Born(1955)]{born1955statistical}
Born, M.
\newblock Statistical interpretation of quantum mechanics.
\newblock {\em Science} {\bf 1955}, {\em 122},~675--679.

\bibitem[Dirac(1938)]{dirac1938classical}
Dirac, P.A.M.
\newblock Classical theory of radiating electrons.
\newblock {\em Proceedings of the Royal Society of London. Series A.
  Mathematical and Physical Sciences} {\bf 1938}, {\em 167},~148--169.

\bibitem[Wheeler and Feynman(1945)]{wheeler1945interaction}
Wheeler, J.A.; Feynman, R.P.
\newblock Interaction with the absorber as the mechanism of radiation.
\newblock {\em Reviews of Modern Physics} {\bf 1945}, {\em 17},~157.

\bibitem[Pegg(1980)]{pegg1980objective}
Pegg, D.
\newblock Objective reality, causality and the Aspect experiment.
\newblock {\em Physics Letters A} {\bf 1980}, {\em 78},~233--234.

\bibitem[Cramer(1980)]{cramer1980generalized}
Cramer, J.G.
\newblock Generalized absorber theory and the Einstein-Podolsky-Rosen paradox.
\newblock {\em Physical Review D} {\bf 1980}, {\em 22},~362.

\bibitem[Harrigan and Spekkens(2010)]{harrigan2010einstein}
Harrigan, N.; Spekkens, R.W.
\newblock Einstein, incompleteness, and the epistemic view of quantum states.
\newblock {\em Foundations of Physics} {\bf 2010}, {\em 40},~125--157.

\bibitem[Arntzenius(1994)]{arntzenius1994spacelike}
Arntzenius, F.
\newblock Spacelike connections.
\newblock {\em The British journal for the philosophy of science} {\bf 1994},
  {\em 45},~201--217.

\bibitem[Maudlin(1996)]{maudlin1996space}
Maudlin, T.
\newblock Space-time in the quantum world. In {\em Bohmian mechanics and
  quantum theory: An appraisal}; Springer,  1996; pp. 285--307.

\bibitem[Wood and Spekkens(2015)]{wood2015lesson}
Wood, C.J.; Spekkens, R.W.
\newblock The lesson of causal discovery algorithms for quantum correlations:
  Causal explanations of Bell-inequality violations require fine-tuning.
\newblock {\em New Journal of Physics} {\bf 2015}, {\em 17},~033002.

\bibitem[Allen \em{et~al.}(2017)Allen, Barrett, Horsman, Lee, and
  Spekkens]{Allen2021QuantumPhysRevX.7.031021}
Allen, J.M.A.; Barrett, J.; Horsman, D.C.; Lee, C.M.; Spekkens, R.W.
\newblock Quantum Common Causes and Quantum Causal Models.
\newblock {\em Phys. Rev. X} {\bf 2017}, {\em 7},~031021.
\newblock
  doi:{\changeurlcolor{black}\href{https://doi.org/10.1103/PhysRevX.7.031021}{\detokenize{10.1103/PhysRevX.7.031021}}}.

\bibitem[Berkovitz(2002)]{berkovitz2002causal}
Berkovitz, J.
\newblock On causal loops in the quantum realm. In {\em Non-locality and
  Modality}; Springer,  2002; pp. 235--257.

\bibitem[DeWitt and Graham(1973)]{dewitt1973many}
DeWitt, B.S.; Graham, N., Eds.
\newblock {\em The many worlds interpretation of quantum mechanics}; Princeton
  University Press,  1973.

\bibitem[Bohm(1952)]{BohmPhysRev.85.166}
Bohm, D.
\newblock A Suggested Interpretation of the Quantum Theory in Terms of "Hidden"
  Variables. I.
\newblock {\em Phys. Rev.} {\bf 1952}, {\em 85},~166--179.
\newblock
  doi:{\changeurlcolor{black}\href{https://doi.org/10.1103/PhysReview85.166}{\detokenize{10.1103/PhysReview85.166}}}.

\bibitem[Pearle(1976)]{Pearle1976PhysRevD.13.857}
Pearle, P.
\newblock Reduction of the state vector by a nonlinear Schr\"odinger equation.
\newblock {\em Phys. Rev. D} {\bf 1976}, {\em 13},~857--868.
\newblock
  doi:{\changeurlcolor{black}\href{https://doi.org/10.1103/PhysRevD.13.857}{\detokenize{10.1103/PhysRevD.13.857}}}.

\bibitem[Ghirardi \em{et~al.}(1986)Ghirardi, Rimini, and
  Weber]{ghirardi1986unified}
Ghirardi, G.C.; Rimini, A.; Weber, T.
\newblock Unified dynamics for microscopic and macroscopic systems.
\newblock {\em Phys. Rev. D} {\bf 1986}, {\em 34},~470.

\bibitem[Reid and Drummond(2021)]{Reid2021}
Reid, M.D.; Drummond, P.D.
\newblock Retrocausal fields, the measurement problem and nonlocality (in
  preparation).

\bibitem[Reid(2017)]{reid2017interpreting}
Reid, M.D.
\newblock Interpreting the macroscopic pointer by analysing the elements of
  reality of a Schr{\"o}dinger cat.
\newblock {\em Journal of Physics A: Mathematical and Theoretical} {\bf 2017},
  {\em 50},~41LT01.

\bibitem[Englert and Brout(1964)]{englert1964broken}
Englert, F.; Brout, R.
\newblock Broken symmetry and the mass of gauge vector mesons.
\newblock {\em Physical Review Letters} {\bf 1964}, {\em 13},~321.

\bibitem[Higgs(1964)]{higgs1964broken}
Higgs, P.W.
\newblock Broken symmetries and the masses of gauge bosons.
\newblock {\em Physical Review Letters} {\bf 1964}, {\em 13},~508.

\bibitem[Guralnik \em{et~al.}(1964)Guralnik, Hagen, and
  Kibble]{guralnik1964global}
Guralnik, G.S.; Hagen, C.R.; Kibble, T.W.
\newblock Global conservation laws and massless particles.
\newblock {\em Physical Review Letters} {\bf 1964}, {\em 13},~585.

\bibitem[Corney and
  Drummond(2006{\natexlab{a}})]{Corney_PD_JPA_2006_GR_fermions}
Corney, J.F.; Drummond, P.D.
\newblock {Gaussian operator bases for correlated fermions}.
\newblock {\em Journal of Physics A} {\bf 2006}, {\em 39},~269--297.

\bibitem[Corney and
  Drummond(2006{\natexlab{b}})]{Corney_PD_PRB_2006_GPSR_fermions}
Corney, J.F.; Drummond, P.D.
\newblock {Gaussian phase-space representations for fermions}.
\newblock {\em Phys. Rev. B} {\bf 2006}, {\em 73},~125112.

\bibitem[Corney and Drummond(2004)]{Corney_PD_PRL2004_GQMC_ferm_bos}
Corney, J.F.; Drummond, P.D.
\newblock {Gaussian Quantum Monte Carlo Methods for Fermions and Bosons}.
\newblock {\em Phys. Rev. Lett.} {\bf 2004}, {\em 93},~260401.

\bibitem[Rosales-Z{\'a}rate and Drummond(2013)]{ResUnityFGO:2013}
Rosales-Z{\'a}rate, L.E.C.; Drummond, P.D.
\newblock {Resolution of unity for fermionic Gaussian operators}.
\newblock {\em Journal of Physics A} {\bf 2013}, {\em 46},~275203.

\bibitem[Bohr(1928)]{Bohr1928Quantum}
Bohr, N.
\newblock The Quantum Postulate and the Recent Development of Atomic Theory.
\newblock {\em Nature} {\bf 1928}, {\em 121},~580--590.
\newblock
  doi:{\changeurlcolor{black}\href{https://doi.org/10.1038/121580a0}{\detokenize{10.1038/121580a0}}}.

\bibitem[Bohr(1935)]{Bohr1935CanQuant}
Bohr, N.
\newblock Can Quantum-Mechanical Description of Physical Reality be Considered
  Complete?
\newblock {\em Phys. Rev.} {\bf 1935}, {\em 48},~696--702.
\newblock
  doi:{\changeurlcolor{black}\href{https://doi.org/10.1103/PhysRev.48.696}{\detokenize{10.1103/PhysRev.48.696}}}.

\bibitem[Friederich(2021)]{Friederich2021Introducing}
Friederich, S.
\newblock Introducing the Q-based interpretation of quantum mechanics.

\bibitem[Thenabadu \em{et~al.}(2020)Thenabadu, Cheng, Pham, Drummond,
  Rosales-Z{\'a}rate, and Reid]{thenabadu2020testing}
Thenabadu, M.; Cheng, G.L.; Pham, T.L.H.; Drummond, L.V.; Rosales-Z{\'a}rate,
  L.; Reid, M.D.
\newblock Testing macroscopic local realism using local nonlinear dynamics and
  time settings.
\newblock {\em Physical Review A} {\bf 2020}, {\em 102},~022202.

\bibitem[Leggett and Garg(1985)]{leggett1985quantum}
Leggett, A.J.; Garg, A.
\newblock Quantum mechanics versus macroscopic realism: Is the flux there when
  nobody looks?
\newblock {\em Phys. Rev. Lett.} {\bf 1985}, {\em 54},~857.

\bibitem[Emary \em{et~al.}(2013)Emary, Lambert, and Nori]{emary2013leggett}
Emary, C.; Lambert, N.; Nori, F.
\newblock Leggett--garg inequalities.
\newblock {\em Reports on Progress in Physics} {\bf 2013}, {\em 77},~016001.

\bibitem[Thenabadu and Reid(2020)]{thenabadu2020bipartite}
Thenabadu, M.; Reid, M.D.
\newblock Bipartite Leggett-Garg and macroscopic Bell inequality violations
  using cat states: distinguishing weak and deterministic macroscopic realism.
\newblock {\em arXiv preprint arXiv:2012.14997} {\bf 2020}.

\bibitem[Yurke and Stoler(1986)]{yurke1986generating}
Yurke, B.; Stoler, D.
\newblock Generating quantum mechanical superpositions of macroscopically
  distinguishable states via amplitude dispersion.
\newblock {\em Physical review letters} {\bf 1986}, {\em 57},~13.

\bibitem[Cartan(1926)]{cartan1926classe}
Cartan, {\'E}.
\newblock Sur une classe remarquable d'espaces de Riemann.
\newblock {\em Bulletin de la Soci{\'e}t{\'e} Math{\'e}matique de France} {\bf
  1926}, {\em 54},~214--264.

\bibitem[Cartan(1935)]{cartan1935domaines}
Cartan, {\'E}.
\newblock Sur les domaines born{\'e}s homogenes de l'espace de n variables
  complexes.
\newblock {\em Abh. Math. Sem. Univ. Hamburg} {\bf 1935}, {\em 11},~116--162.

\bibitem[Perelomov(1972)]{perelomov1972coherent}
Perelomov, A.M.
\newblock Coherent states for arbitrary Lie group.
\newblock {\em Communications in Mathematical Physics} {\bf 1972}, {\em
  26},~222--236.

\bibitem[Arecchi \em{et~al.}(1972)Arecchi, Courtens, Gilmore, and
  Thomas]{Arecchi1972}
Arecchi, F.; Courtens, E.; Gilmore, R.; Thomas, H.
\newblock {Atomic Coherent States in Quantum Optics}.
\newblock {\em Phys. Rev. A} {\bf 1972}, {\em 6},~2211--2237.
\newblock
  doi:{\changeurlcolor{black}\href{https://doi.org/10.1103/PhysRevA.6.2211}{\detokenize{10.1103/PhysRevA.6.2211}}}.

\bibitem[Gilmore \em{et~al.}(1975)Gilmore, Bowden, and Narducci]{Gilmore1975}
Gilmore, R.; Bowden, C.; Narducci, L.
\newblock {Classical-quantum correspondence for multilevel systems}.
\newblock {\em Phys. Rev. A} {\bf 1975}, {\em 12},~1019--1031.
\newblock
  doi:{\changeurlcolor{black}\href{https://doi.org/10.1103/PhysRevA.12.1019}{\detokenize{10.1103/PhysRevA.12.1019}}}.

\bibitem[Zhang \em{et~al.}(1990)Zhang, Feng, and Gilmore]{Zhang1990}
Zhang, W.M.; Feng, D.H.; Gilmore, R.
\newblock {Coherent states: Theory and some applications}.
\newblock {\em Reviews of Modern Physics} {\bf 1990}, {\em 62},~867--927.
\newblock
  doi:{\changeurlcolor{black}\href{https://doi.org/10.1103/RevModPhysical62.867}{\detokenize{10.1103/RevModPhysical62.867}}}.

\bibitem[Schwarzschild(1903)]{schwarzschild1903elektrodynamik}
Schwarzschild, K.
\newblock Zur Elektrodynamik. II. die elementare elektrodynamische Kraft.
\newblock {\em Nachrichten von der Gesellschaft der Wissenschaften zu
  G{\"o}ttingen, Mathematisch-Physikalische Klasse} {\bf 1903}, {\em
  1903},~132--141.

\bibitem[Tetrode(1922)]{tetrode1922causal}
Tetrode, H.M.
\newblock On the causal connection of the world, an extension of classical
  dynamics.
\newblock {\em Zeitschrift fur Physik} {\bf 1922}, {\em 10},~317--328.

\bibitem[Fokker(1929)]{fokker1929invarianter}
Fokker, A.D.
\newblock Ein invarianter Variationssatz f{\"u}r die Bewegung mehrerer
  elektrischer Massenteilchen.
\newblock {\em Zeitschrift f{\"u}r Physik} {\bf 1929}, {\em 58},~386--393.

\bibitem[Hoyle and Narlikar(1971)]{hoyle1971electrodynamics}
Hoyle, F.; Narlikar, J.
\newblock Electrodynamics of direct interparticle action: II. Relativistic
  treatment of radiative processes.
\newblock {\em Annals of Physics} {\bf 1971}, {\em 62},~44--97.

\bibitem[Cramer(1986)]{cramer1986transactional}
Cramer, J.G.
\newblock The transactional interpretation of quantum mechanics.
\newblock {\em Reviews of Modern Physics} {\bf 1986}, {\em 58},~647.

\bibitem[Kastner(2006)]{kastner2006cramer}
Kastner, R.E.
\newblock Cramer's transactional interpretation and causal loop problems.
\newblock {\em Synthese} {\bf 2006}, {\em 150},~1--14.

\bibitem[Price(2008)]{price2008toy}
Price, H.
\newblock Toy models for retrocausality.
\newblock {\em Studies in History and Philosophy of Science Part B: Studies in
  History and Philosophy of Modern Physics} {\bf 2008}, {\em 39},~752--761.

\bibitem[Wharton and Argaman(2020)]{wharton2020colloquium}
Wharton, K.B.; Argaman, N.
\newblock Colloquium: Bell's theorem and locally mediated reformulations of
  quantum mechanics.
\newblock {\em Reviews of Modern Physics} {\bf 2020}, {\em 92},~021002.

\bibitem[Shrapnel and Costa(2018)]{Shrapnel2018causationdoesnot}
Shrapnel, S.; Costa, F.
\newblock Causation does not explain contextuality.
\newblock {\em {Quantum}} {\bf 2018}, {\em 2},~63.
\newblock
  doi:{\changeurlcolor{black}\href{https://doi.org/10.22331/q-2018-05-18-63}{\detokenize{10.22331/q-2018-05-18-63}}}.

\bibitem[Stratonovich(1966)]{stratonovich1971probability}
Stratonovich, R.
\newblock A New Representation for Stochastic Integrals and Equations.
\newblock {\em SIAM J. Control} {\bf 1966}, {\em 4},~362.

\bibitem[Graham(1977{\natexlab{a}})]{Graham1977Covariant}
Graham, R.
\newblock Covariant formulation of non-equilibrium statistical thermodynamics.
\newblock {\em Zeitschrift f{\"u}r Physik B} {\bf 1977}, {\em 26},~397--405.
\newblock
  doi:{\changeurlcolor{black}\href{https://doi.org/10.1007/BF01570750}{\detokenize{10.1007/BF01570750}}}.

\bibitem[Graham(1977{\natexlab{b}})]{graham1977lagrangian}
Graham, R.
\newblock Lagrangian for diffusion in curved phase space.
\newblock {\em Phys. Rev. Lett.} {\bf 1977}, {\em 38},~51.

\bibitem[Bell(1964)]{Bell1964}
Bell, J.S.
\newblock {On the Einstein-Podolsky-Rosen paradox}.
\newblock {\em Physics} {\bf 1964}, {\em 1},~195--200.

\bibitem[Pawula(1967)]{pawula1967approximation}
Pawula, R.F.
\newblock Approximation of the linear Boltzmann equation by the Fokker-Planck
  equation.
\newblock {\em Physical Review} {\bf 1967}, {\em 162},~186.

\bibitem[Tomonaga(1946)]{Tomonaga1946Relativistic}
Tomonaga, S.
\newblock {On a Relativistically Invariant Formulation of the Quantum Theory of
  Wave Fields}.
\newblock {\em Progress of Theoretical Physics} {\bf 1946}, {\em 1},~27--42,
  \href{http://xxx.lanl.gov/abs/https://academic.oup.com/ptp/article-pdf/1/2/27/24027031/1-2-27.pdf}{{\normalfont
  [https://academic.oup.com/ptp/article-pdf/1/2/27/24027031/1-2-27.pdf]}}.
\newblock
  doi:{\changeurlcolor{black}\href{https://doi.org/10.1143/PTP.1.27}{\detokenize{10.1143/PTP.1.27}}}.

\bibitem[Feynman(1948)]{Feynman1948RelativisticPhysRev.74.1430}
Feynman, R.P.
\newblock Relativistic Cut-Off for Quantum Electrodynamics.
\newblock {\em Phys. Rev.} {\bf 1948}, {\em 74},~1430--1438.
\newblock
  doi:{\changeurlcolor{black}\href{https://doi.org/10.1103/PhysRev.74.1430}{\detokenize{10.1103/PhysRev.74.1430}}}.

\bibitem[Schwinger(1948)]{Schwinger1948QuantumPhysRev.74.1439}
Schwinger, J.
\newblock Quantum Electrodynamics. I. A Covariant Formulation.
\newblock {\em Phys. Rev.} {\bf 1948}, {\em 74},~1439--1461.
\newblock
  doi:{\changeurlcolor{black}\href{https://doi.org/10.1103/PhysRev.74.1439}{\detokenize{10.1103/PhysRev.74.1439}}}.

\bibitem[Schwinger(1949)]{Schwinger1949QuantumPhysRev.75.651}
Schwinger, J.
\newblock Quantum Electrodynamics. II. Vacuum Polarization and Self-Energy.
\newblock {\em Phys. Rev.} {\bf 1949}, {\em 75},~651--679.
\newblock
  doi:{\changeurlcolor{black}\href{https://doi.org/10.1103/PhysRev.75.651}{\detokenize{10.1103/PhysRev.75.651}}}.

\bibitem['t~Hooft(1971)]{thooft1971renormalizable}
't~Hooft, G.
\newblock Renormalizable lagrangians for massive Yang-Mills fields.
\newblock {\em Nuclear physics: B} {\bf 1971}, {\em 35},~167--188.

\bibitem['t~Hooft(1972)]{tHooftveltman1972regularization}
't~Hooft, G. amd~Veltman, M.
\newblock Regularization and renormalization of gauge fields.
\newblock {\em Nuclear Physics B} {\bf 1972}, {\em 44},~189--213.

\bibitem[Kibble(1980)]{kibble1980some}
Kibble, T.W.
\newblock Some implications of a cosmological phase transition.
\newblock {\em Physics Reports} {\bf 1980}, {\em 67},~183--199.

\bibitem[Zee(2010)]{zee2010quantum}
Zee, A.
\newblock {\em Quantum field theory in a nutshell}; Vol.~7, Princeton
  university press,  2010.

\bibitem[Kuhn(1957)]{kuhn1957copernican}
Kuhn, T.S.
\newblock {\em The Copernican revolution: Planetary astronomy in the
  development of western thought}; Vol.~16, Harvard University Press,  1957.

\bibitem[Popper(1989)]{popper1989logik}
Popper, K.R.
\newblock {\em Logik der Forschung}; JCB Mohr T{\"u}bingen,  1989.

\bibitem[Gomis and Weinberg(1996)]{gomis1996nonrenormalizable}
Gomis, J.; Weinberg, S.
\newblock Are nonrenormalizable gauge theories renormalizable?
\newblock {\em Nuclear Physics B} {\bf 1996}, {\em 469},~473--487.

\bibitem[Isham(1993)]{isham1993canonical}
Isham, C.J.
\newblock Canonical quantum gravity and the problem of time. In {\em Integrable
  systems, quantum groups, and quantum field theories}; Springer,  1993; pp.
  157--287.

\bibitem[Bohr(1996)]{bohr1996discussion}
Bohr, N.
\newblock Discussion with Einstein on epistemological problems in atomic
  physics. In {\em Niels Bohr Collected Works}; Elsevier,  1996; Vol.~7, pp.
  339--381.

\bibitem[Reid \em{et~al.}(2009)Reid, Drummond, Bowen, Cavalcanti, Lam, Bachor,
  Andersen, and Leuchs]{Reid:2009_RMP81}
Reid, M.D.; Drummond, P.D.; Bowen, W.P.; Cavalcanti, E.G.; Lam, P.K.; Bachor,
  H.A.; Andersen, U.L.; Leuchs, G.
\newblock \textit{Colloquium}: The Einstein-Podolsky-Rosen paradox: From
  concepts to applications.
\newblock {\em Reviews of Modern Physics} {\bf 2009}, {\em 81},~1727--1751.

\bibitem[Einstein(1949)]{einstein1949albert}
Einstein, A.
\newblock Albert Einstein: Autobiographical Notes. In {\em Albert Einstein -
  Philosopher Scientist}; Schilpp, P.A., Ed.; Library of Living Philosophers,
  Open Court Publishing, Chicago, USA,  1949; Vol.~7, pp. 2--95.

\bibitem[Dieks(1988)]{dieks1988formalism}
Dieks, D.
\newblock The formalism of quantum theory: an objective description of reality?
\newblock {\em Annalen der Physik} {\bf 1988}, {\em 500},~174--190.

\bibitem[Stapp(2001)]{Stapp2001Quantum}
Stapp, H.P.
\newblock Quantum Theory and the Role of Mind in Nature.
\newblock {\em Foundations of Physics} {\bf 2001}, {\em 31},~1465--1499.
\newblock
  doi:{\changeurlcolor{black}\href{https://doi.org/10.1023/A:1012682413597}{\detokenize{10.1023/A:1012682413597}}}.

\bibitem[Von~Neumann(2018)]{von2018mathematical}
Von~Neumann, J.
\newblock {\em Mathematical foundations of quantum mechanics: New edition};
  Princeton university press,  2018.

\bibitem[Bub(2010)]{bub2010neumann}
Bub, J.
\newblock Von Neumann's `no hidden variables' proof: A re-appraisal.
\newblock {\em Foundations of Physics} {\bf 2010}, {\em 40},~1333--1340.

\bibitem[Deckert \em{et~al.}(2019)Deckert, Esfeld, and
  Oldofredi]{deckert2019persistent}
Deckert, D.A.; Esfeld, M.; Oldofredi, A.
\newblock A persistent particle ontology for quantum field theory in terms of
  the Dirac sea.
\newblock {\em The British Journal for the Philosophy of Science} {\bf 2019},
  {\em 70},~747--770.

\bibitem[Bell(2004)]{bell2004speakable}
Bell, J.S.
\newblock {\em Speakable and unspeakable in quantum mechanics: Collected papers
  on quantum philosophy}; Cambridge University Press,  2004.

\bibitem[Everett(1957)]{Everett:1957}
Everett, H.
\newblock {"Relative State" Formulation of Quantum Mechanics}.
\newblock {\em Reviews of Modern Physics} {\bf 1957}, {\em 29},~454--462.

\bibitem[Zeh(1970)]{zeh1970interpretation}
Zeh, H.D.
\newblock On the interpretation of measurement in quantum theory.
\newblock {\em Foundations of Physics} {\bf 1970}, {\em 1},~69--76.

\bibitem[Griffiths(1984)]{griffiths1984consistent}
Griffiths, R.B.
\newblock Consistent histories and the interpretation of quantum mechanics.
\newblock {\em J. Stat. Phys.} {\bf 1984}, {\em 36},~219--272.

\bibitem[Omn{\`e}s(1992)]{omnes1992consistent}
Omn{\`e}s, R.
\newblock Consistent interpretations of quantum mechanics.
\newblock {\em Reviews of Modern Physics} {\bf 1992}, {\em 64},~339.

\bibitem[Gell-Mann and Hartle(1993)]{gell1993classical}
Gell-Mann, M.; Hartle, J.B.
\newblock Classical equations for quantum systems.
\newblock {\em Phys. Rev. D} {\bf 1993}, {\em 47},~3345.

\bibitem[Omn{\`e}s(1999)]{omnes1999understanding}
Omn{\`e}s, R.
\newblock {\em Understanding quantum mechanics}; Princeton University Press,
  1999.

\bibitem[Van~Fraassen(1974)]{fraassen1974einstein}
Van~Fraassen, B.C.
\newblock The Einstein-Podolsky-Rosen Paradox.
\newblock {\em Synthese} {\bf 1974}, {\em 29},~291.

\bibitem[Dowker and Kent(1995)]{Dowker1995PropertiesPhysRevLett.75.3038}
Dowker, F.; Kent, A.
\newblock Properties of Consistent Histories.
\newblock {\em Phys. Rev. Lett.} {\bf 1995}, {\em 75},~3038--3041.
\newblock
  doi:{\changeurlcolor{black}\href{https://doi.org/10.1103/PhysRevLett.75.3038}{\detokenize{10.1103/PhysRevLett.75.3038}}}.

\bibitem[Nelson(1966)]{nelson1966derivation}
Nelson, E.
\newblock Derivation of the Schr{\"o}dinger equation from Newtonian mechanics.
\newblock {\em Phys. Rev.} {\bf 1966}, {\em 150},~1079.

\bibitem[Parisi \em{et~al.}(1981)Parisi, Wu, et~al.]{parisi1981perturbation}
Parisi, G.; Wu, Y.S.; others.
\newblock Perturbation theory without gauge fixing.
\newblock {\em Scientia Sinica} {\bf 1981}, {\em 24},~483--469.

\bibitem[Budiyono and Rohrlich(2017)]{budiyono2017quantum}
Budiyono, A.; Rohrlich, D.
\newblock Quantum mechanics as classical statistical mechanics with an ontic
  extension and an epistemic restriction.
\newblock {\em Nature Communications} {\bf 2017}, {\em 8},~1306.

\bibitem[Grabert \em{et~al.}(1979)Grabert, H{\"a}nggi, and
  Talkner]{grabert1979quantum}
Grabert, H.; H{\"a}nggi, P.; Talkner, P.
\newblock Is quantum mechanics equivalent to a classical stochastic process?
\newblock {\em Phys. Rev. A} {\bf 1979}, {\em 19},~2440.

\bibitem[Nelson(2012)]{nelson2012review}
Nelson, E.
\newblock Review of stochastic mechanics.
\newblock  Journal of Physics: Conference Series. IOP Publishing,  2012, Vol.
  361, p. 012011.

\bibitem[Blanchard \em{et~al.}(1986)Blanchard, Golin, and
  Serva]{blanchard1986repeated}
Blanchard, P.; Golin, S.; Serva, M.
\newblock Repeated measurements in stochastic mechanics.
\newblock {\em Physical Review D} {\bf 1986}, {\em 34},~3732.

\bibitem[Damgaard and H{\"u}ffel(1987)]{damgaard1987stochastic}
Damgaard, P.H.; H{\"u}ffel, H.
\newblock Stochastic quantization.
\newblock {\em Physics Reports} {\bf 1987}, {\em 152},~227--398.

\end{thebibliography}
\end{document}